\def\Journal#1#2#3#4{{#1} {\bf #2}, #3 (#4)}
\def\bibitem#1{\noindent{\bf#1}.}
\def\cite#1{$^{\bf #1}$}

\def\PRL{\it Phys. Rev. Lett.}
\def\PRD{{\it Phys. Rev.} D}


\def\APJ{{\it Ap. J.}}
\def\APJL{{\it Ap. J. Lett.}}


\def\mpl{M_{Pl}}
\def\gev{{\rm GeV}}
\def\lamp{\Lambda_p}
\def\lamh{\Lambda_h}
\def\nut{\tilde\nu}
\def\log{{\rm log}}
\def\zte{\zeta (E)}
\def\pscmsr{{\rm cm^{-2}\,s^{-1}\,sr^{-1}}}
\def\pscm{{\rm cm^{-2}\,s^{-1}}}
\def\cms{{\rm cm^3\,s^{-1}}}

\def\vavsq{\langle v^2\rangle}
\def\mfs{M_{fs}}

\def\mz{M_0}
\def\omz{\Omega_0}
\def\hz{h_0}
\def\tz{T_0}
\def\tf{T_f}
\def\tk{T_K}
\def\deg{\,^\circ{\rm K}}
\def\sv{\langle\sigma v\rangle}
\def\vc{V_c}
\def\mathnew{\mathsurround=0pt}
\def\simov#1#2{\lower .5pt\vbox{\baselineskip0pt \lineskip-.5pt
		\ialign{$\mathnew#1\hfil##\hfil$\crcr#2\crcr\sim\crcr}}}
\def\simless{\mathrel{\mathpalette\simov <}}
\def\simgreat{\mathrel{\mathpalette\simov >}}
\def\kms{\,{\rm km\,s^{-1}}}
\def\cm{\,{\rm cm}}
\def\au{\,{\rm A.U.}}
\def\pc{\,{\rm pc}}
\def\kpc{\,{\rm kpc}}
\def\mpc{\,{\rm Mpc}}
\def\msun{M_\odot}
\def\yr{\,{\rm yr}}
\def\rhocl{\rho_{cl}}
\def\rhoz{\rho_0}
\def\rhobar{\overline{\rho}}
\def\delcl{\delta_{cl}}
\def\delclav{\langle\delcl\rangle}

\centerline{DENSITY FLUCTUATIONS IN THE GALACTIC HALO}
\centerline{AND EXPERIMENTAL SEARCHES FOR DARK MATTER}
\bigskip
\centerline{ Ira Wasserman }
\medskip
\centerline{Center for Radiophysics and Space Research}
\centerline{Cornell University, Ithaca, NY 14853 USA}

\medskip
\centerline{\sl Talk Presented at Sakharov Conference on Physics, May 1996}
\bigskip

\noindent
Clumping of elementary dark matter in the Galaxy halo may be
inevitable. If so, the nondetection of certain dark matter 
candidates could simply mean that the local halo density is low.
Conversely, indirect detection of annihilation products could
be facilitated, perhaps to an embarrassing degree.
\bigskip
The interpretation of the results of experiments designed to search for 
elementary particle dark matter in the Milky Way halo relies on specific 
suppositions concerning the local mass density and velocity distribution
of the unseen particles. Predicted detection rates\cite{1} have all 
been computed
under the assumptions that the halo density profile 
is smooth, and the velocity
distribution is at least approximately Maxwellian, with a one-dimensional
velocity dispersion $\sigma=\vc/\sqrt{2}\approx 160\kms$, where
$\vc\approx 220\kms$ is the Galactic rotation speed. However, an experiment
that operates for a time $T\sim 1\yr$ only probes a region that extends to
a distance $D\sim\sigma T\sim 5\times 10^{14}\cm\approx 35\au$, which is 
about the distance to the planet Pluto; for comparison, the distance to 
the center of the Galaxy is 
$D_0\approx 8.5\kpc\approx 2\times 10^9\au$. While
few would doubt that the halo density may be regarded as smooth on some
intermediate length scale, say $\simless 0.01-1\kpc$, 
there is reason to question
whether the halo really ought to be perfectly smooth on length scales of order
the size of the inner Solar System, $\sim 10^{-8}-10^{-6}\kpc$.

If the halo is clumpy on small scales, then it is possible that presently
the Earth is not immersed in a bath of dark matter particles, so that the
non-detection of such particles need not imply their absence. As a simple
but extreme example, suppose that the dark matter 
particles are organized into
lumps of internal density $\rhocl\gg\rhobar$, where 
$\rhobar\approx 0.01\msun\pc^{-3}$
is the mean halo density locally; then the 
probability that the Earth sits in
a dark matter clump is the volume filling factor, $f=\rhobar/\rhocl\ll 1$.
If instead the halo contains a continuous mass spectrum of dark matter lumps,
so that $\rhobar \psi (M) dM$ is the 
contribution to the local halo density from
clumps of mass $M$ and internal density $\rhocl (M)$, 
then the volume filling 
factor of the halo is
$$
f=\rhobar\int{dM\,\psi (M) \rhocl^{-1}(M)}, \eqno(1)
$$
which may also be very small if the dark matter is concentrated into dense
clumps. Consequently, experimentalists searching for elementary particle
dark matter in a clumpy halo could simply be unlucky. Conversely, a
fortunate experimenter bathed in a dark matter cloud could detect halo
particles at an incredibly large rate compared to predictions based
on the assumption of a smooth halo.

Although the problem of small scale halo structure is generic, we focus 
here on a specific case: neutralinos of mass $m\simgreat\gev$
whose annihilation is predominantly
s-wave.\cite{2} The cosmology of such
particles is governed by $m$, and a dimensionless 
parameter, $C=\mpl m\sv$, where $\mpl$ is the Planck mass and $\sv$ is the
annihilation rate coefficient for the fermions. Annihilation equilibrium
fails at a temperature $\tf\sim m/\ln C$, allowing a significant present
day abundance, amounting to a total mass density $\rhoz\sim m\tz^3\ln C/C$,
where $\tz$ is an appropriate present day temperature (e.g. about $2\deg$
for neutrinos); at density parameter $\omz$ and Hubble constant
$H_0=100\hz\kms\mpc^{-1}$, $C\sim 10^9m(\gev)(\omz\hz^2)^{-1}$.
The neutralinos remain in thermal contact with the ambient gas of
annihilation products until the temperature falls to $\tk\sim mC^{-1/4}$,
after which they decouple completely from the rest of the cosmological
soup.\cite{3}

After $\tk$, the neutralinos stream freely along straight line 
paths. Consequently, any fluctuations in the density of dark matter
particles will be erased on scales smaller than the free-streaming
length, which corresponds to a mass\cite{3}
$\mfs\sim(\mpl^3/m^2)\ln C/C^{5/8}
\sim 10^{-4}\msun(\omz\hz^2)^{5/8}/[m(\gev)]^{21/8}$. 
Moreover, by Liouville's theorem, the microscopic phase space density
is preserved after $\tk$, when its value was $\lamp\sim\ln C/C^{11/8}
\sim 10^{-11}[\omz\hz^2/m(\gev)]^{11/8}$.
The coarse-grained phase space density of the halo is
$\lamh\sim 10^{-32}/[m(\gev)]^4\ll\lamp$, which suggests that some
fine-grained clustering of neutralinos in the halo is inevitable.

The nature of the clustering is still not certain. One possibility 
is that dark matter clumps of mass $\mfs$, which are expected to be
among the first objects to condense cosmologically, survive intact
to the present day and dominate the Galactic halo. Research along
these general lines has been reported at this conference by Gurevich
and collaborators, who also suggest that non-baryonic clumps may be
the MACHOs. Another possibility is that clumps are not permanent but
come and go in a dynamical fashion as the result of small scale
`gravitational turbulence' in the Galaxy halo: locally space-time
dependent density fluctuations with possibly stationary statistical
properties.

A simple picture that could lead to a stationary statistical state
for the halo would involve binary fragmentation and merger of dark matter
clumps in the halo. To be concrete, let us focus on a discrete model
in which $N_k$ is the number of clumps of mass $M_k=\mz 2^{-k}$, where
$k$ is an integer; in a continuous model, $N_k$ would correspond
to $\psi (M)$ in Eq. (1). Suppose that fragmentation
of a cloud of mass $M_k$ results in the production of a pair of bound
clouds of mass $M_{k+1}$; merger of a pair of clouds with equal masses
$M_k$ forms a cloud of mass $M_{k-1}$. The net rate of formation of
clouds of mass $M_k$ is then
$$
{dN_k\over dt}=-(\nu_k+2\nut_{k-1})N_k+2\nu_{k-1}N_{k-1}+\nut_kN_{k+1},
\eqno(2)
$$
where $\nu_k$ is the rate of fragmentation of a mass $M_k$ and $\nut_k$
is the rate of merger to form a mass $M_k$. If we assume that $\nu_k
\sim\sqrt{G\rho_k}$ and $\nut_k=c\nu_k$, where $\rho_k$ is the internal
density for mass $M_k$ and $c$ is independent of scale, then 
Eq. (2) has the steady state solution $N_k\propto M_k^{\log_2c}$;
notice that this is independent of the dependence of density on mass
scale, and that the mass spectrum is tilted toward small masses for 
$c<1/2$. (This steady-state solution only requires the assumption that
$\nut_c=c\nu_k$, not the additional 
physical conjecture that the characteristic
merger and collapse rates are approximately free-fall.)
Assuming that the basic physics of collapse and merger is
scale free also leads to $\rho_k\propto M_k^{-2\eta}$; if fragmentation
tends to leave behind bound pairs, then $\eta<1$ in general.\cite{4}
The mass
spectrum extends down to the `Jeans mass', where the internal phase
space density of a lump is $\sim\lamp$; this is generally below $\msun$
but above $\mfs$ for $m\simgreat 1\gev$. The picture outlined above
can be generalized to a continuous mass spectrum, and may also apply
if fragmentation is replaced by formation of low mass `satellites' in
the course of a merger (i.e. if mergers produce, in addition to a large
clump comprising most of the colliding mass, smaller ejected sub-lumps),
although the value of $\eta$ could differ in that case.
Needless to say, other pictures for the development of fluctuations might
be more compelling than the one presented here, but may or may not
lead to stationary statistics of the fluctuations.

Although {\it direct} detection of elementary particle dark matter would
be frustrated by clumpiness in the halo, {\it indirect} detection would
be facilitated, perhaps to an embarrassing extent. Let us suppose that
annihilation of dark matter particles produces some observable product
-- e.g. high energy gamma rays -- with an efficiency $\zte dE/E$ for
energies near $E$. Then the (number) flux of observable annihilation
products from a single clump of mass $M$ a distance $D$ from Earth
is (using $\sv\sim C/\mpl m\approx 10^{-27}(\omz\hz^2)^{-1}\cms$)
$$
E{dF_{cl}\over dE}\sim{\sv\rhocl (M)M\zte\over 4\pi D^2m^2}
\sim 3\times 10^{-9}\pscm{\delcl\zte M/\msun\over
\omz\hz^2[m(\gev)]^2[D(\pc)]^2}
\eqno(3)
$$
where $\delcl=\rhocl/\rhobar$, and the integrated flux from
the halo of our Galaxy is 
$$
E{dF_G\over d\Omega dE}\sim
{\sv\rhobar^2D_0\delclav\zte\over 4\pi m^2}
\sim 3\times 10^{-7}{\pscmsr}{\delclav\zte
\over\omz\hz^2[m(\gev)]^2}
\eqno(4)
$$
where $\delclav$ is the mean clump density contrast weighted
by $\psi (M)$. (For predominantly p-wave annihilation, $\sv
\propto\vavsq\simless 10^{-6}$ and Eqs. (3)
and (4) would be correspondingly lower; the 
admixture of s- and p-wave in $\sv$ is model-dependent.)
For comparison,\cite{5} the excess gamma ray flux
above 100 MeV measured by EGRET that is not correlated with
Galactic HI column density is $I_{ex}=1.47\pm 0.03\times 10^{-5}
\pscmsr$, and the isotropic component of this excess is
estimated to be $I_{iso}=1.5\pm 0.3\times 10^{-5}\pscmsr$.
Since the spectrum of the diffuse emission falls with increasing
energy,\cite{6} a halo flux as large as indicated by Eq. (4)
could be excessive at photon energies $\simgreat 1\gev$
unless $\delclav\zte$ is small. Also, EGRET has 
apparently found a small number of sources unidentified with
any previously known astronomical objects.\cite{7}
These sources have
total fluxes $\sim 10^{-7}\pscm$ above 100 MeV, and no larger
than $\sim 10^{-8}\pscm$ above 1 GeV. Although such fluxes could
be consistent with Eq. (3) (depending on the value
of $\delcl\zte/D^2$) it is probably unlikely that annihilation
of multi-GeV particles would produce sub-GeV photons 
preferentially,\cite{8} and the mysterious sources could have a
more mundane origin than annihilation in neutralino clumps.
To the extent that small-scale concentrations of dark matter 
may be shown to be
required physically, the absence of detectable annihilation products
could be used to argue against the existence of certain neutralino
candidates.
\medskip
This research was supported in part by NSF grants AST 91-19475
and AST 93-15375, and NASA grants NAG 5-3097 and NAG 5-2762.
\bigskip
\noindent{\bf References}
\medskip
\bibitem{1} See, for example, M. Goodman and E. Witten, 
\Journal{\PRD}{31}{3059}{1985},
I. Wasserman, \Journal{\PRD}{33}{2071}{1986}, 
K. Griest, \Journal{\PRD}{38}{2357}{1988}.

\bibitem{2} B.W. Lee and S. Weinberg, \Journal{\PRL}{39}{165}{1977}.

\bibitem{3} J.E. Gunn {\it et al}, \Journal{\APJ}{223}{1015}{1978}.

\bibitem{4} W.I. Newman and I. Wasserman, \Journal{\APJ}{354}{411}{1990}.

\bibitem{5} A. Chen, J. Dwyer and P. Kaaret, \Journal{\APJL}{445}{L109}{1995}.

\bibitem{6} C.E. Fichtel, G.A. Simpson and D.J. Thompson, 
\Journal{\APJ}{222}{833}{1978}.

\bibitem{7} P. Sreekumar {\it et al.}, \Journal{\APJ}{464}{628}{1996}.

\bibitem{8} See, for example, J. Silk and M. Srednicki, 
\Journal{\PRL}{53}{624}{1984},
for a discussion of gamma ray production via halo photino annihilation.

\end

\end{document}